\documentclass[reprint,aps,prd,nofootinbib,preprintnumbers,twocolumns,showpacs,10pt, superscriptaddress]{revtex4-1}
\usepackage{epsfig}
\usepackage{epstopdf}
\usepackage{hyperref}
\usepackage{amssymb}
\usepackage{amsmath}
\usepackage{amsfonts}
\usepackage{graphicx}
\usepackage{bm}
\usepackage[usenames,dvipsnames]{color}
\hypersetup{
  colorlinks   = true, %Colours links instead of ugly boxes
  urlcolor     = blue, %Colour for external hyperlinks
  linkcolor    = blue, %Colour of internal links
  citecolor    = red   %Colour of citations
}
\def	\be	{\begin{equation}}
\def	\ee	{\end{equation}}
\def	\bqt	{\begin{quote}}
\def	\eqt	{\end{quote}}

\begin{document}

\title{New wave equation for ultrarelativistic particles}

\author{Gin\'{e}s R.P\'{e}rez Teruel}

\affiliation{Departamento de F\'{i}sica Te\'{o}rica, Universidad de Valencia, Burjassot-46100, Valencia, Spain} 

\begin{abstract}
\begin{center}
{\bf Abstract}
\end{center}
\noindent
Starting from first principles and general assumptions based on the energy-momentum relation of the Special Theory of Relativity we present a novel wave equation for ultrarelativistic matter. This wave equation arises when particles satisfy the condition, $p>>m$, i.e, when the energy-momentum relation can be approximated by, $E\simeq p+\frac{m^{2}}{2p}$. Interestingly enough, such as the Dirac equation, it is found that this wave equation includes spin in a natural way. Furthermore, the free solutions of this wave equation contain plane waves that are completely equivalent to those of the theory of neutrino oscillations. Therefore, the theory reproduces some standard results of the Dirac theory in the limit $p>>m$, but offers the possibility of an explicit Lorentz Invariance Violation of order, $\mathcal{O}((mc)^{4}/p^{2})$. As a result, the theory could be useful to test small departures from Dirac equation and Lorentz Invariance at very high energies. On the other hand, the wave equation can also describe particles of spin 1 by a simple substitution of the spin operators, $\boldsymbol{\sigma}\rightarrow\boldsymbol{\alpha}$. In addition, it naturally admits a Lagrangian formulation and a Hamiltonian formalism. We also discuss the associated conservation laws that arise through the symmetry transformations of the Lagrangian.
\end{abstract}

\maketitle
\section{Introduction}
\label{Introduction} 
\thispagestyle{empty}

\noindent
The Dirac equation is one of the most beautiful creations of human intellect. It opened a new era in particle physics, providing impressive results among which we can mention the prediction of antiparticles such as the positron. Indeed, the prediction of the positron by Dirac in 1928, --and the posterior discovery by Anderson in 1932--, marks one of the major events in the history of science. However, in spite of their impressive agreement with experiments, it is not clear whether the Dirac equation will \emph{always} prevail as an untouchable scientific truth. We specifically refer to the ultra-high energy regimes, i.e, $p>>m$ , where possible deviations from Lorentz invariance might be present. These possible violations of Lorentz invariance have received attention in the literature and are being extensively studied in the last years, specially in the context of ultra-high energy cosmic-rays.\cite{Macc,Bie}\\

It is not the purpose of this letter the study of a competitor to the Dirac wave equation. There is no doubt that fermions follow the Dirac equation, whose unrivalled success in the understanding of matter is not subjected to discussion. However, fermions such as the electron also satisfy the Schr\"odinger equation, but in the low-energy limit. Indeed, as is well known, the Dirac equation replaces the Schr\"odinger equation in the relativistic regime. Then, it seems natural to wonder: Is there life beyond the Dirac equation? In this work we try to provide an answer to this question. In particular, we want to explore if a theoretical alternative to the Dirac equation at very high energies can be possible. From a theoretical point of view, the study of such a possibility is not only a legitimate research program, but also a quite interesting intellectual exercise. 

Indeed, as we will see, in the regime $p>>m$, it seems to be room for the existence of another matter wave equation, which can be derived starting from first principles. This matter wave equation also includes spin in a natural way. In fact, we will show that the wave equation is not only suitable for fermions of spin 1/2, but does also describe massive bosons of spin 1 by means of a substitution of the spin operators $\boldsymbol{\sigma}\rightarrow\boldsymbol{\alpha}$, where $\boldsymbol{\alpha}$ are the Majorana-Oppenheimer matrices \cite{Maj,Opp31}.

The paper is organized as follows. In sec.\ref{Derivation}, we derive the wave equation starting from first principles. In sec.\ref{Canonical formulation}, we establish the Lagrangian formulation and the Hamiltonian formalism. Finally, in sec.\ref{conclusions}, we present the conclusions of this work..
\section{Derivation and physical interpretation}
\label{Derivation} 
\thispagestyle{empty}

\noindent
In order to derive a wave equation in Physics, we have to focus on the energy-momentum relation assumed. In the same fashion that Schr\"odinger's equation is derived assuming a classical relation, $E=p^{2}/2m$, the Klein-Gordon equation can be obtained taking the relativistic, $E^{2}=m^{2}+p^{2}$, and the Dirac equation emerges assuming a linear relation, $E=\alpha^{i}p_{i}+\beta m$. Then, it seems natural to wonder what wave equation would correspond to the case, $E=p+\frac{m^{2}}{2p}$, which is the subject of this work.
In the theoretical discussion that follows, we will maintain the constants $\hbar$, $c$, in all the expressions unless otherwise noted. Let us begin the discussion with the energy-momentum relation of the Special Theory of Relativity (STR) for a free particle with positive energy
\be\label{Energy}
E=\sqrt{c^{2}p^{2}+m^{2}c^{4}}
\, 
\ee
Where $m$ is the rest mass. As is well known, in the non-relativistic limit, $p<<mc$, we can approximate this equation as
\be\label{Energy2}
E\simeq mc^{2}\Big(1+\frac{p^{2}}{2m^{2}c^{2}}\Big)= mc^{2}+\frac{p^{2}}{2m}
\,
\ee
However, in the ultra-relativistic limit, $mc << p$, equation (\ref{Energy}) provides
\be\label{Energy3}
E\simeq cp\Big(1+\frac{m^{2}c^{2}}{2p^{2}}\Big)=cp+\frac{m^{2}c^{3}}{2p}
\,
\ee
Multiplying by $p$ the last equation and making explicit the operator representation, $\widehat{E}\displaystyle\rightarrow i\hbar\frac{\partial}{\partial t}$,  $\widehat{p}\displaystyle\rightarrow-i\hbar\nabla$,  we arrive to the following partial differential wave equation

\be\label{Wave_equation}
|\nabla|\frac{\partial\psi}{\partial t}=-c\nabla^{2}\psi+\frac{m^{2}c^{3}}{2\hbar^{2}}\psi
\,
\ee
\\
Where, $\nabla^{2}\equiv\partial^{2}_{x}+\partial^{2}_{y}+\partial^{2}_{z}$ is the Laplacian operator and, $|\nabla|\equiv(\partial^{2}_{x}+\partial^{2}_{y}+\partial^{2}_{z})^{1/2}$ its square root. We can express the square root of the Laplacian as, $|\nabla|=\vec{\sigma}\cdot\nabla$; In cartesian coordinates 
\begin{equation}\label{spinors}
|\nabla|=\vec{\sigma}\cdot\nabla=\sigma_{x}\partial_{x}+\sigma_{y}\partial_{y}+\sigma_{z}\partial_{z}
\,
\end{equation}
Where, $\sigma_{x}$, $\sigma_{y}$, $\sigma_{z}$ are certain operators. The condition, $(\vec{\sigma}\cdot\nabla)^{2}=\nabla^{2}$ can be written as
\begin{equation}
\Big(\sigma_{x}\partial_{x}+\sigma_{y}\partial_{y}+\sigma_{z}\partial_{z}\Big)\cdot \Big(\sigma_{x}\partial_{x}+\sigma_{y}\partial_{y}+\sigma_{z}\partial_{z}\Big)=\partial^{2}_{x}+\partial^{2}_{y}+\partial^{2}_{z}
\,
\end{equation}
Then, the operators $\sigma_{i}$ are subjected to the constraints
\begin{eqnarray}
\sigma^{2}_{i}=I
\qquad
\{\sigma_{i},\sigma_{j}\}=2\delta_{ij}I
\end{eqnarray}

This allows to choose a representation of the sigmas given in terms of the Pauli matrices

\begin{eqnarray}\label{Pauli}
\sigma_{x}=\begin{pmatrix} 0 & 1 \\ 1 & 0 \end{pmatrix} \qquad
\sigma_{y} = \begin{pmatrix} 0 & -i \\ i & 0 \end{pmatrix} \qquad \sigma_{z} = \begin{pmatrix} 1 & 0 \\ 0 & -1 
\end{pmatrix}
\end{eqnarray}

Finally, we can write the explicit form of the wave equation (\ref{Wave_equation}) is terms of the Pauli matrices as follows

\begin{equation}\label{wave_equationexplicit}
\sigma_{x}\frac{\partial^{2}\psi}{\partial x\partial t}+\sigma_{y}\frac{\partial^{2}\psi}{\partial y\partial t}+\sigma_{z}\frac{\partial^{2}\psi}{\partial z\partial t}=-c\nabla^{2}\psi+\frac{m^{2}c^{3}}{2\hbar^{2}}\psi
\end{equation}
\\
Or in the compact form
\begin{equation}\label{compact}
(\vec{\sigma}\cdot\nabla)\frac{\partial \psi}{\partial t}=-c\nabla^{2}\psi+\frac{m^{2}c^{3}}{2\hbar^{2}}\psi
\,
\end{equation}
This explicit emergence of the Pauli matrices (\ref{Pauli}) in the wave equation indicates that (\ref{compact}) is describing particles of spin $1/2$ in the ultra-high energy regime, $p>>mc$.
It is important to note that this wave equation is not covariant (with respect to a Lorentz Transformation). This is because the energy and the momentum do not share the same power in the energy-momentum relation, which implies that the spatial and temporal derivatives that appear in the wave equation are not of the same order. Since we are neglecting in the power series expansion of (\ref{Energy3}) all the terms beyond, $m^{2}c^{2}/2p$, the breakdown of Lorentz invariance (LI), given by (\ref{Energy3}) is therefore of order, $\mathcal{O}((mc)^{4}/p^{2})$, this represents a very small violation of LI, almost negligible.

At any rate, experiments have the last word. If a wave equation explains the experimental results but fails to be Lorentz invariant, then the problem is not necessarily in the wave equation. In this sense, it is well known that there is no clue so far of Lorentz invariance violations (LIV) in particle physics experiments. However, a large amount of theoretical effort is being spent trying to study this possibility, specially in the context of the cosmic rays, which are conformed by ultra-relativistic particles.\cite{Macc,Bie}. Although the wave equation is not LI, it is interesting to note that with some manipulations it can be written in a form that resembles a covariant equation:
\\
\begin{equation}\label{covariant}
(\sigma_{\mu}\partial^{\mu})\frac{\partial\psi}{\partial x^{0}}=\Big(-\partial_{\mu}\partial^{\mu}+\frac{m^{2}c^{2}}{2\hbar^{2}}\Big)\psi
\end{equation}

where $\sigma_{0}=I_{2x2}$. Indeed, since $\sigma_{0}\partial^{0}\partial_{0}\psi=\frac{1}{c^{2}}\frac{\partial^{2}\psi}{\partial t^{2}}I_{2x2}$, this equation leads to
\\
\begin{equation}\label{covariant2}
\frac{1}{c^{2}}\frac{\partial^{2}\psi}{\partial t^{2}}+(\boldsymbol{\sigma}\cdot\nabla)\frac{\partial\psi}{c\partial t}=\Big(-\nabla^{2}+\frac{1}{c^{2}}\frac{\partial^{2}}{\partial t^{2}}+\frac{m^{2}c^{2}}{2\hbar^{2}}\Big)\psi
\end{equation}

Then, we have a cancellation between the second-order partial derivatives with respect to the time coordinate, recovering (\ref{compact}). From the observation of (\ref{covariant2}), we can see that the term $\partial_{0}\psi$ is preventing the covariance of the wave equation. Indeed, this is the zero component of a four-vector, $\partial_{\mu}\psi$, referred to a particular coordinate system, meanwhile all the other terms of the wave equation can be written in a manifest Lorentz invariant fashion. On the other hand, the Hamiltonian form of the wave equation can easily be provided:
\begin{equation}\label{covariant}
(-i\sigma_{\mu}\partial^{\mu})\mathcal{\hat{H}}\psi=\Big(-\hbar\partial_{\mu}\partial^{\mu}+\frac{m^{2}c^{2}}{2\hbar}\Big)\psi
\end{equation}
\\
where
\begin{equation}\label{standard Hamiltonian}
\mathcal{\hat{H}}\psi=i\hbar\frac{\partial\psi}{\partial t}
\end{equation}
$\mathcal{\hat{H}}$ is the standard Hamiltonian (or time-evolution) operator. In the last part of this work we shall see a peculiar aspect of this theory, namely, that the time evolution operator $\mathcal{\hat{H}}$ and the canonical Hamiltonian $\hat{H_{c}}$ that comes from the Lagrangian formulation are not the same algebraic object.
\subsection{The plane wave solution. Dispersion relation, phase and group velocities of the ultrarelativistic waves.} 
Given the wave equation (\ref{wave_equationexplicit}), it seems natural to look for a solution with a plane-wave structure

\be\label{plane_wave}
\psi(r,t)={\chi \choose \phi}e^{i(\vec{k}\cdot\vec{r}-wt)}
\,
\ee
\\
Where, $\vec{k}\cdot\vec{r}=k_{x}x+k_{y}y+k_{z}z$. Substituing this ansatz in (\ref{wave_equationexplicit}) we obtain after a bit of algebra the following matrix equation

\begin{equation}\label{Matrix}
\begin{pmatrix}
k_{z}\omega-ck^{2}-\frac{m^{2}c^{3}}{2\hbar^{2}}& (k_{x}-ik_{y})\omega\\
(k_{x}+ik_{y})\omega&-k_{z}\omega-ck^{2}-\frac{m^{2}c^{3}}{2\hbar^{2}}\\
\end{pmatrix}
{\chi \choose \phi}={0 \choose 0}
\end{equation}
\\
Where, $k^{2}=k_{x}^{2}+k_{y}^{2}+k_{z}^{2}$. In order to obtain non-trivial solutions we must impose the condition, $\det\hat{A}=0$.
It gives the result
\begin{equation}
\det\hat{A}=-k_{z}^{2}\omega^{2}+\Big(ck^{2}+\frac{m^{2}c^{3}}{2\hbar^{2}}\Big)^{2}-\omega^{2}(k_{x}^{2}+k_{y}^{2})=0
\end{equation}

Which implies a dispersion relation, $w(k)$ given by

\begin{equation}\label{dispersion}
w(k)=\pm\Big(ck+\frac{m^{2}c^{3}}{2\hbar^{2}k}\Big)
\end{equation}
The minus sign corresponds to the negative energy solutions.Then, taking into account both possibilities, the most general solution of the wave equation will be a superposition of positive and negative energy modes, namely:
\begin{equation}\label{general solution}
\psi(x)=\sum_{k}\Big(a(k)\boldsymbol{u}(k)e^{i(\vec{k}\cdot\vec{r}-\omega t)}+b^{\dagger}(k)\boldsymbol{v}(k)e^{-i(\vec{k}\cdot\vec{r}-\omega t)}\Big)
\end{equation}

 with an spinor of positive energy $\boldsymbol{u}(k)$, and another of negative energy, $\boldsymbol{v}(k)$ satisfying the orthogonality condition $\boldsymbol{u}(k)\cdot\boldsymbol{v}(k)=0$. Indeed, in order to better understand the last result, let us consider the particular case of the one-dimensional propagation along the z-axis. The simplified version of (\ref{wave_equationexplicit}) will be
\be\label{Wave_equation2}
\sigma_{z}\frac{\partial^{2}\psi}{\partial z\partial t}=-c\frac{\partial^{2}\psi}{\partial z^{2}}+\frac{m^{2}c^{3}}{2\hbar^{2}}\psi
\,
\ee
To solve this wave equation we take an ansatz similar to (\ref{plane_wave})
\begin{equation}\label{plane_wave2}
\psi(z,t)={\chi \choose \phi}e^{i(kz-wt)}
\,
\end{equation}
For simplicity we have denoted the $z$ component of the wave vector $\vec{k}$ simply as $k$. The substitution of (\ref{plane_wave2}) in (\ref{Wave_equation2}) provides the following system 

\begin{eqnarray}
\frac{kw}{c}\chi=\Big(k^{2}+\frac{m^{2}c^{2}}{2\hbar^{2}}\Big)\chi\,\\
\frac{kw}{c}\phi=-\Big(k^{2}+\frac{m^{2}c^{2}}{2\hbar^{2}}\Big)\phi\,
\,
\end{eqnarray}
These relations imply 
\begin{eqnarray}\label{dispersion_relation}
w_{\chi}(k)=ck+\frac{m^{2}c^{3}}{2\hbar^{2}k}\,\\
w_{\phi}(k)=-\Big(ck+\frac{m^{2}c^{3}}{2\hbar^{2}k}\Big)\,
\,
\end{eqnarray}
Of course, they are nothing else than the dispersion relations that we deduced for the 3D case in (\ref{dispersion}). Note that the field, $\phi$ that appears in (\ref{plane_wave}-\ref{plane_wave2}) is the piece of the wave function that corresponds to the negative energy solution. i.e, the field associated to the antiparticle.
 
On the other hand, it is possible to compute directly from (\ref{dispersion}) the phase and group velocities associated to the positive energy solutions.
\be\label{phase_velocity}
V_{ph}=\frac{w}{k}=c+\frac{m^{2}c^{3}}{2\hbar^{2}k^{2}}=c\Big(1+\Big(\frac{mc}{\sqrt{2}\hbar k}\Big)^{2}\Big)
\,
\ee
\be\label{group_velocity}
V_{g}=\frac{dw}{dk}=c-\frac{m^{2}c^{3}}{2\hbar^{2}k^{2}}=c\Big(1-\Big(\frac{mc}{\sqrt{2}\hbar k}\Big)^{2}\Big)
\,
\ee
Note that in the previous relations $mc<<\hbar k$. Then, although the phase velocity of the ultra-relativistic wave satisfies, $V_{ph}\geq c$, the group velocity (the meaningful concept related with the true energy propagation of the wave), cannot exceed the speed of light. We can therefore conclude that the propagation of these waves is causal and consistent with the STR, the superluminal propagation is not possible in this theory.
Furthermore, given the values of $V_{g}$ and $V_{ph}$, we can assure after a straightforward computation that their product has an upper bound given by $c^{2}$

\begin{equation}
V_{ph}\cdot V_{g}=c^{2}\Big(1-\Big(\frac{mc}{\sqrt{2}\hbar k}\Big)^{4}\Big)
\,
\end{equation}

On the other hand, note that by means of the dispersion relation (\ref{dispersion_relation}), the plane wave solution with positive energy of (\ref{plane_wave2}), can be written in the following manner
\begin{align}\label{neutrino_waves}
\chi(z,t)&=\displaystyle\chi(0)\exp\Big(i(kz-w_{\chi}t)\Big)\nonumber\\
&=\displaystyle\chi(0)\exp\Big(i(kz-kct-\frac{m^{2}c^{3}}{2\hbar^{2}k}t)\Big)\nonumber\\
&\simeq\chi(0)\exp\Big(-i\frac{m^{2}c^{2}z}{2\hbar^{2}k}\Big)
\,
\end{align}
Where we can approximate $z\simeq ct$ if the ultra-relativistic particle travels close to the speed of light.
The wave function (\ref{neutrino_waves}) is a standard result frequently found within the context of the theory of neutrino oscillations\cite{Bil,Bala} . This theory makes the initial assumption that the mass eigenfunctions that describe the propagation of such particles are plane-waves, $|\nu_{i}(z,t)\rangle = \exp(i(k_{i}z-w_{i}t))|\nu_{i}(0)\rangle$, then it is used the approximation (in natural units), $E=p+m^{2}/2p$ to simplify the argument of the exponential and finally obtain 
\begin{equation}\label{neutrino_waves2} 
 |\nu_{i}(z)\rangle = \exp\Big(-i \frac{m_{i}^{2} z}{2p}\Big)|\nu_{i}(0)\rangle
\,
\end{equation}
\\
Where $z$ is the distance between the neutrino production and detection points. Note that both wave functions have the same structure. It is worth noting that we have been able to derive this standard result following a non-standard approach. Indeed, we have proved that the family of plane waves (\ref{neutrino_waves2}), are only particular solutions of the ultra-relativistic wave equation (\ref{Wave_equation}).\\
On the other hand, eigenstates with different masses propagate at different speeds, this is evident following equation (\ref{group_velocity}) which establishes the dependence of the group velocity upon $m^{2}$. This fact is also directly derived from the wave equation.\\
\subsection{ The spin 1 case. Majorana-Oppenheimer matrices}
Let us briefly illustrate how the formalism can be naturally adapted to describe ultrarelativistic bosons of spin 1. This can be achieved by a subtle change of operators without changing the structure itself of the wave equation. Indeed, let us rewrite (\ref{compact}),
in the following form:

\begin{equation}\label{compact2}
(\boldsymbol{\alpha}\cdot\nabla)\frac{\partial \boldsymbol{\psi}}{\partial t}=-c\nabla^{2}\boldsymbol{\psi}+\frac{m^{2}c^{3}}{2\hbar^{2}}\boldsymbol{\psi}
\,
\end{equation}
\\
 
The difference with respect to the spin 1/2 case lies in the left hand side, but it is worth noting that the structure of the wave equation remains the same with the only modification $\boldsymbol{\sigma}\rightarrow\boldsymbol{\alpha}$. This means that instead of the Pauli matrices, now we have another spin operators. These operators are given by the following matrices:

\begin{eqnarray}\label{Majorana-Oppenheimer1}
\alpha^{1}=\begin{pmatrix} 0 & 0 & 0 \\ 0 & 0 & i \\ 0 & -i & 0 \end{pmatrix} \qquad
\alpha^{2} = \begin{pmatrix} 0 & 0 & -i \\ 0 & 0 & 0\\i & 0 & 0 \end{pmatrix}\qquad
\end{eqnarray}
and
\begin{eqnarray}\label{Majorana-Oppenheimer2}
\alpha^{3} = \begin{pmatrix} 0 & i & 0 \\ -i & 0 &0\\0 & 0 & 0\end{pmatrix}
\end{eqnarray}

These matrices satisfy the angular--momentum commutation rules
\begin{equation}\label{commutation}
[\alpha_i,\alpha_k]=-i\varepsilon_{ikl}\alpha_l\quad
\end{equation}

Matrices (\ref{Majorana-Oppenheimer1},\ref{Majorana-Oppenheimer2}) were introduced by Majorana \cite{Maj} and Oppenheimer \cite{Opp31} in their independent attempt to formulate Maxwell's Electrodynamics as the Field Theory of a massless spin 1 particle within the framework of a Dirac-type equation. Such as the Pauli matrices, these operators are hermitian $\boldsymbol{\alpha}^{\dagger}=\boldsymbol{\alpha}$, and obey the commutation rules of the rotation group SO(3) (\ref{commutation}). Since they are 3x3 matrices, the field $\boldsymbol{\psi}$ must be decomposed as $\boldsymbol{\psi}=(\psi_1,\psi_2,\psi_3)$. In particular, the plane wave solution will be of the form:
\begin{equation}\label{plane_wave_spin1}
\boldsymbol{\psi}(r,t)=\left(\begin{array}{c}\psi_{1}\\\psi_{2}\\\psi_{3}\end{array}\right)e^{i(\vec{k}\cdot\vec{r}-\omega t)}
\,
\end{equation}
\\
The substitution of this ansatz in the wave equation (\ref{compact2}) provides, after some elementary manipulations, a homogeneous matrix system $\hat{A}\boldsymbol{\psi}=\boldsymbol{0}$, similar to that of the spin 1/2 case (\ref{Matrix}), but $\hat{A}$ will be now a 3x3 hermitian matrix. Indeed, the explicit form of this homogeneous system is the following:
\small
\begin{equation}
\begin{pmatrix}
-ck^{2}-\frac{m^{2}c^{3}}{2\hbar^{2}}& ik_{z}\omega & -ik_{y}\omega\\
-ik_{z}\omega & -ck^{2}-\frac{m^{2}c^{3}}{2\hbar^{2}} &ik_{x}\omega\\
ik_{y}\omega& -ik_{x}\omega & -ck^{2}-\frac{m^{2}c^{3}}{2\hbar^{2}}\\
\end{pmatrix}
\left(\begin{array}{c}\psi_{1}\\\psi_{2}\\\psi_{3}\end{array}\right)=\left(\begin{array}{c}0\\0\\0\end{array}\right)
\end{equation}
\normalsize
\\
Note that $\hat{A}^{\dagger}=\hat{A}$. To obtain non-trivial solutions, we must impose again the consistency condition $\det\hat{A}=0$. Through this condition one gets the result:
\begin{equation}
\Big(ck^{2}+\frac{m^{2}c^{3}}{2\hbar^{2}}\Big)\Big(k^{2}\omega^{2}-\Big(ck^{2}+\frac{m^{2}c^{3}}{2\hbar^{2}}\Big)^{2}\Big)=0
\end{equation}
which implies

\begin{equation}
\omega(k)=\pm\Big(ck+\frac{m^{2}c^{3}}{2\hbar^{2}k}\Big)
\end{equation}

This is the dispersion relation $\omega(k)$ expected for a ultrarelativistic particle of energy-momentum relation $E\simeq p+m^{2}/2p$, a result that was already derived for the spin 1/2 case, in (\ref{dispersion}) and (\ref{dispersion_relation}). As in the spin 1/2 case, the negative frequency solution (negative energy), corresponds to the antiparticle.

Finally, the Hamiltonian form of the wave equation for the spin 1 case in a ``covariant" fashion can be written as :
\begin{equation}
(-i\alpha_{\mu}\partial^{\mu})\mathcal{\hat{H}}\boldsymbol{\psi}=\Big(-\hbar\partial_{\mu}\partial^{\mu}+\frac{m^{2}c^{2}}{2\hbar}\Big)\boldsymbol{\psi}
\end{equation}
where $\alpha_{0}=I_{3x3}$
\newpage
\section{Canonical formulation}
\label{Canonical formulation}
\thispagestyle{empty}

\noindent
In this section we proceed to the construction of the Lagrangian and Hamiltonian formulation of the theory. As is well known, a remarkable feature of Field Theory is that all the well defined matter wave equations can be derived from a Lagrangian density, from which a continuity equation, $\frac{\partial\rho}{\partial t}+\nabla\cdot\boldsymbol{J}=0$, follows. The aim of this section is to prove that the wave equation (\ref{compact}) also admits naturally a Lagrangian formalism. The canonical analysis is a powerful tool, not only to study the symmetry transformations of the Lagrangian which allows to apply Noether's theorem to collect the associated conservation laws, but also to build the associated Hamiltonian, a necessary step to carry out the canonical quantization of the field.\footnote{The study of the canonical quantization of the field will be the subject of future work}
\subsection{Lagrangian formulation} 
The starting point of the canonical analysis are the following Lagrangians:

\begin{equation}\label{Lagrangian density}
\mathcal{L}_{1/2}=\frac{m^{2}c^{2}}{2\hbar^{2}}\psi\psi^{\dagger}+\nabla\psi\cdot\nabla \psi^{\dagger}+\frac{1}{2c}\Big(\frac{\partial\psi}{\partial t}\boldsymbol{\sigma}\cdot\nabla\psi^{\dagger}+\frac{\partial \psi^{\dagger}}{\partial t}\boldsymbol{\sigma}\cdot\nabla\psi\Big) 
\,
\end{equation}

\begin{equation}\label{Lagrangian density_bosons}
\mathcal{L}_{1}=\frac{m^{2}c^{2}}{2\hbar^{2}}\boldsymbol{\psi}\boldsymbol{\psi}^{\dagger}+\nabla\boldsymbol{\psi}\cdot\nabla \boldsymbol{\psi}^{\dagger}+\frac{1}{2c}\Big(\frac{\partial\boldsymbol{\psi}}{\partial t}\boldsymbol{\alpha}\cdot\nabla\boldsymbol{\psi}^{\dagger}+\frac{\partial \boldsymbol{\psi}^{\dagger}}{\partial t}\boldsymbol{\alpha}\cdot\nabla\boldsymbol{\psi}\Big) 
\,
\end{equation}
where $c$ is the speed of light. As is well known, the matrices $\boldsymbol{\sigma}$, $\boldsymbol{\alpha}$, are hermitian, $\sigma_{i}^{\dagger}=\sigma_{i}$, $\alpha_{i}^{\dagger}=\alpha_{i}$ which guarantees the hermiticity of both Lagrangians, $\mathcal{L}=\mathcal{L}^{\dagger}$. Since the spin 1 case is identical to the 1/2 case with the replacement $\boldsymbol\sigma\rightarrow\boldsymbol{\alpha}$ in the Lagrangian, we shall restrict ourselves to the study of the 1/2 case, and it should be understood that a similar analysis holds for the case of spin 1. Having made this clarification, the Euler-Lagrange (E-L) equations for the fields $\psi$, $\psi^{\dagger}$, are given by

\begin{equation}\label{Lagrangian density}
\mathcal{L}=\frac{m^{2}c^{2}}{2\hbar^{2}}\psi\psi^{\dagger}+\nabla\psi\cdot\nabla \psi^{\dagger}+\frac{1}{2c}\Big(\frac{\partial\psi}{\partial t}\vec{\sigma}\cdot\nabla\psi^{\dagger}+\frac{\partial \psi^{\dagger}}{\partial t}\vec{\sigma}\cdot\nabla\psi\Big) 
\,
\end{equation}
\\
Where $c$ is the speed of light. As is well known, the Pauli matrices are hermitian, $\sigma_{i}^{\dagger}=\sigma_{i}$, which guarantees the hermiticity of the Lagrangian, $\mathcal{L}=\mathcal{L}^{\dagger}$. The Euler-Lagrange (E-L) equations for the fields $\psi$, $\psi^{\dagger}$, are given by

\begin{align}\label{Euler-Lagrange}
\partial_{\mu}\Big[\frac{\partial\mathcal{L}}{\partial(\partial_{\mu}\psi)}\Big]-\frac{\partial\mathcal{L}}{\partial\psi}=0,
\qquad
\partial_{\mu}\Big[\frac{\partial\mathcal{L}}{\partial(\partial_{\mu}\psi^{\dagger})}\Big]-\frac{\partial\mathcal{L}}{\partial\psi^{\dagger}}=0
\end{align}
\\
Making explicit the summation over the index $\mu$, the E-L equation associated to the hermitian field $\psi^{\dagger}$ will be
\begin{equation}\label{Euler-Lagrange2}
\partial_{0}\Big[\frac{\partial\mathcal{L}}{\partial(\partial_{0}\psi^{\dagger})}\Big]+\nabla\cdot\Big[\frac{\partial\mathcal{L}}{\partial(\nabla\psi^{\dagger})}\Big]-\frac{\partial\mathcal{L}}{\partial\psi^{\dagger}}=0
\,
\end{equation}
\\
Applying the derivatives of (\ref{Euler-Lagrange2}) to the Lagrangian density (\ref{Lagrangian density}), we find
\begin{align}\label{Euler-Lagrange3}
\frac{\partial\mathcal{L}}{\partial(\partial_{0}\psi^{\dagger})}=\frac{1}{2c}\vec{\sigma}\cdot\nabla\psi,
\qquad
\frac{\partial\mathcal{L}}{\partial(\nabla\psi^{\dagger})}=\frac{1}{2c}\frac{\partial\psi}{\partial t}\vec{\sigma}+\nabla\psi
\end{align}
\\
The substitution of these results in (\ref{Euler-Lagrange2}) gives
\begin{align}
0&=\partial_{0}\Big[\frac{\partial\mathcal{L}}{\partial(\partial_{0}\psi^{\dagger})}\Big]+\nabla\cdot\Big[\frac{\partial\mathcal{L}}{\partial(\nabla\psi^{\dagger})}\Big]-\frac{\partial\mathcal{L}}{\partial\psi^{\dagger}}\nonumber\\
&=\frac{\partial}{\partial t}\Big(\frac{1}{2c}\vec{\sigma}\cdot\nabla\psi\Big)+\nabla\cdot\Big(\frac{1}{2c}\frac{\partial\psi}{\partial t}\vec{\sigma}+\nabla\psi\Big)-\frac{m^{2}c^{2}}{2\hbar^{2}}\psi\nonumber\\
&=\frac{1}{c}(\vec{\sigma}\cdot\nabla)\frac{\partial\psi}{\partial t}+\nabla^{2}\psi-\frac{m^{2}c^{2}}{2\hbar^{2}}\psi
\end{align}
\\
We have therefore been able to derive the wave equation (\ref{compact}) from a Lagrangian density by means of the corresponding E-L equations.
\subsection{Global gauge invariance. Noether's theorem and conserved current}
The Lagrangian density (\ref{Lagrangian density}) is invariant under the transformation
\begin{equation}\label{gauge}
\psi\rightarrow\psi^{\prime}=e^{i\theta}\psi
\,
\end{equation}
Then, according to Noether's theorem it must exist a conserved quantity. Indeed, it can be proved that the associated current $J^{\mu}$, satisfies the differential equation $\partial_{\mu}J^{\mu}=0$, where
\begin{equation}\label{current}
J^{\mu}=\Big[\frac{\partial\mathcal{L}}{\partial(\partial_{\mu}\psi)}\Big]\delta\psi+\Big[\frac{\partial\mathcal{L}}{\partial(\partial_{\mu}\psi^{\dagger})}\Big]\delta\psi^{\dagger}
\,
\end{equation}
\\
This result implies that there exists a certain ``charge" $Q\equiv\int_{V} J^{0}d^{3}x$, which is a constant of motion, i.e, $dQ/dt=0$. For a transformation of the type given by (\ref{gauge}) we have
\begin{align}
\delta\psi=\psi^{\prime}-\psi=\Big(e^{i\theta}-1\Big)\psi\approx i\theta\psi\nonumber\\
\delta\psi^{\dagger}\approx-i\theta\psi^{\dagger}
\,
\end{align}
\\
Substituing the results of equations (\ref{Euler-Lagrange3}) together with these last identities in (\ref{current}), we obtain
\begin{align}
J^{0}&=\Big[\frac{\partial\mathcal{L}}{\partial(\partial_{0}\psi)}\Big]\delta\psi+\Big[\frac{\partial\mathcal{L}}{\partial(\partial_{0}\psi^{\dagger})}\Big]\delta\psi^{\dagger}\nonumber\\
&=\frac{i\theta}{2c}\Big[\Big(\vec{\sigma}\cdot\nabla\psi^{\dagger}\Big)\psi-\Big(\vec{\sigma}\cdot\nabla\psi\Big)\psi^{\dagger}\Big]
\,
\end{align}

\begin{align}
\vec{J}&=\Big[\frac{\partial\mathcal{L}}{\partial(\nabla\psi)}\Big]\delta\psi+\Big[\frac{\partial\mathcal{L}}{\partial(\nabla\psi^{\dagger})}\Big]\delta\psi^{\dagger}\nonumber\\
&=i\theta\Big[\psi\Big(\frac{1}{2c}\frac{\partial\psi^{\dagger}}{\partial t}\vec{\sigma}+\nabla\psi^{\dagger}\Big)-\psi^{\dagger}\Big(\frac{1}{2c}\frac{\partial\psi}{\partial t}\vec{\sigma}+\nabla\psi\Big)\Big]
\,
\end{align}
\\
It is straightforward to show that these functions satisfy the hermiticity condition, $J^{0}=(J^{0})^{\dagger}$, $\vec{J}=\vec{J}^{\dagger}$. On the other hand, since the parameter $\theta$ is an arbitrary constant, we can take $\theta=1$. Finally, the conserved ``charge" will be
\begin{equation}\label{charge}
Q=\displaystyle\int_{V} J^{0}d^{3}x=\frac{i}{2c}\int_{V} \Big[\Big(\vec{\sigma}\cdot\nabla\psi^{\dagger}\Big)\psi-\Big(\vec{\sigma}\cdot\nabla\psi\Big)\psi^{\dagger}\Big]d^{3}x
\end{equation}
\\
In order to verify the robustness of these results, we can check if the divergence $\partial_{\mu}J^{\mu}$ vanishes or not. After a bit of algebra we find
\begin{align}\label{divergence}
\partial_{\mu}J^{\mu}&=\partial_{0}J^{0}+\nabla\cdot\vec{J}\nonumber\\
&=i\psi\Big(\frac{1}{c}\vec{\sigma}\cdot\nabla\frac{\partial\psi^{\dagger}}{\partial t}+\nabla^{2}\psi^{\dagger}\Big)-i\psi^{\dagger}\Big(\frac{1}{c}\vec{\sigma}\cdot\nabla\frac{\partial\psi}{\partial t}+\nabla^{2}\psi\Big)\nonumber\\
&=i\frac{m^{2}c^{2}}{2\hbar^{2}}\psi\psi^{\dagger}-i\frac{m^{2}c^{2}}{2\hbar^{2}}\psi^{\dagger}\psi=0
\,
\end{align}
\\
Therefore, the divergence $\partial_{\mu}J^{\mu}$ vanishes identically as expected. Then, we can conclude that the Lagrangian formulation of the ultra-relativistic wave equation is a consistent theory, and the Lagrangian density (\ref{Lagrangian density}) has a global gauge symmetry compatible with a conserved current.
\subsection{Local gauge invariance}
The generalization to the $U(1)$ case is straightforward. As is well known, for a local phase transformation, i.e, $\psi\rightarrow\psi^{\prime}=e^{i\theta(x)}\psi$, the usual derivative transforms in the following way
\begin{align}
\partial_{\mu}\psi\rightarrow\partial_{\mu}\psi^{\prime}&=\partial_{\mu}(e^{i\theta(x)}\psi)=\partial_{\mu}(e^{i\theta(x)})\psi+e^{i\theta(x)}\partial_{\mu}\psi\nonumber\\
&=e^{i\theta(x)}(i\partial_{\mu}\theta(x))\psi+e^{i\theta(x)}\partial_{\mu}\psi\nonumber\\
&=e^{i\theta(x)}[i\partial_{\mu}\theta(x)+\partial_{\mu}]\psi
\,
\end{align}
Then, the Lagrangian density (\ref{Lagrangian density}) is no longer invariant under this transformation and we must look for a generalization. This generalization is
\begin{equation}\label{Lagrangian_U(1)}
\mathcal{L}=\frac{m^{2}c^{2}}{2\hbar^{2}}\psi\psi^{\dagger}+\mathcal{D}_{i}\psi(\mathcal{D}_{i}\psi)^{\dagger}+\frac{1}{2c}\Big(\mathcal{D}_{0}\psi(\vec{\sigma}\cdot\vec{\mathcal{D}}\psi)^{\dagger}+(\mathcal{D}_{0}\psi)^{\dagger}\vec{\sigma}\cdot\vec{\mathcal{D}}\psi\Big) 
\,
\end{equation}
\\
Indeed, the covariant derivative, $\mathcal{D}_{\mu}\equiv(\partial_{\mu}+A_{\mu})$ is subjected to the transformation rule
\begin{equation}
\mathcal{D}_{\mu}\psi\rightarrow\mathcal{D}^{\prime}_{\mu}\psi^{\prime}=(\partial_{\mu}+A_{\mu}^{\prime})e^{i\theta(x)}\psi=e^{i\theta(x)}[i\partial_{\mu}\theta(x)+\partial_{\mu}+A_{\mu}^{\prime}]\psi
\end{equation}
\\
Then, in order to compensate the term $i\partial_{\mu}\theta(x)$, we take the condition $A_{\mu}\rightarrow A^{\prime}_{\mu}=A_{\mu}-i\partial_{\mu}\theta(x)$. which implies, $\mathcal{D}_{\mu}^{\prime}\psi^{\prime}=e^{i\theta(x)}\mathcal{D}_{\mu}\psi$, assuring the invariance of (\ref{Lagrangian_U(1)})

\subsection{The Hamiltonian formalism}
Armed with a consistent Lagrangian theory, the next logical step after the analysis of the internal transformations such as (\ref{gauge}) is the study of the external symmetries and the Hamiltonian formalism. The canonical energy-momentum tensor that comes from the Lagrangian density (\ref{Lagrangian density}), under space-time translational invariance is
\begin{equation}\label{energy-momentum tensor}
T^{\mu}_{\nu}=\frac{\partial\mathcal{L}}{\partial(\partial_{\mu}\psi)}(\partial_{\nu}\psi)+\frac{\partial\mathcal{L}}{\partial(\partial_{\mu}\psi^{\dagger})}(\partial_{\nu}\psi^{\dagger})-\delta^{\mu}_{\nu}\mathcal{L}\\
\,
\end{equation}

On the other hand, we can define the Hamiltonian density $\mathcal{H}$, as
\begin{equation}\label{Hamiltonian_density}
\mathcal{H}=T^{0}_{0}=\frac{\partial\mathcal{L}}{\partial\dot{\psi}}\dot{\psi}+\frac{\partial\mathcal{L}}{\partial\dot{\psi^{\dagger}}}\dot{\psi^{\dagger}}-\mathcal{L}=\pi(x)\dot{\psi}+\pi^{\dagger}(x)\dot{\psi^{\dagger}}-\mathcal{L}
\,
\end{equation}\\
The canonical momenta, $\pi(x)$, $\pi^{\dagger}(x)$ are given by the following relations
\begin{align}\label{canonical_momenta}
\pi(x)=\frac{\partial\mathcal{L}}{\partial\dot{\psi}}=\frac{1}{2c}\vec{\sigma}\cdot\nabla\psi^{\dagger},
\qquad
\pi^{\dagger}(x)=\frac{\partial\mathcal{L}}{\partial\dot{\psi^{\dagger}}}=\frac{1}{2c}\vec{\sigma}\cdot\nabla\psi
\end{align}\\
Therefore, substituing the last results in (\ref{Hamiltonian_density}) and using (\ref{Lagrangian density}) we obtain
\begin{align}
\mathcal{H}&=\pi\dot{\psi}+\pi^{\dagger}\dot{\psi^{\dagger}}-\mathcal{L}=\frac{1}{2c}\Big(\dot{\psi}\vec{\sigma}\cdot\nabla\psi^{\dagger}+\dot{\psi}^{\dagger}\vec{\sigma}\cdot\nabla\psi\Big)-\mathcal{L}\nonumber\\
&=\frac{1}{2c}\Big(\dot{\psi}\vec{\sigma}\cdot\nabla\psi^{\dagger}+\dot{\psi}^{\dagger}\vec{\sigma}\cdot\nabla\psi\Big)-\frac{m^{2}c^{2}}{2\hbar^{2}}\psi\psi^{\dagger}-\nabla\psi\cdot\nabla \psi^{\dagger}\nonumber\\
&-\frac{1}{2c}\Big(\frac{\partial\psi}{\partial t}\vec{\sigma}\cdot\nabla\psi^{\dagger}+\frac{\partial \psi^{\dagger}}{\partial t}\vec{\sigma}\cdot\nabla\psi\Big)\nonumber\\
&=-\frac{m^{2}c^{2}}{2\hbar^{2}}\psi\psi^{\dagger}-\nabla\psi\cdot\nabla \psi^{\dagger}
\end{align}\\
The conserved currents and their ``charges", such as $J^{0}$ and $T^{0}_{0}$, are only determined up to a constant. This means that we are free to redefine $\mathcal{H}\equiv-T^{0}_{0}$, in order to have a positive defined Hamiltonian density. Then, we can adopt
\begin{align}
\mathcal{H}&=\frac{m^{2}c^{2}}{2\hbar^{2}}\psi\psi^{\dagger}+\nabla\psi\cdot\nabla \psi^{\dagger}\nonumber\\
&=\frac{m^{2}c^{2}}{2\hbar^{2}}\psi\psi^{\dagger}+\nabla\cdot\Big(\psi^{\dagger}\nabla\psi\Big)-\psi^{\dagger}\nabla^{2}\psi
\,
\end{align}
With this result, the relation between the Hamiltonian $H$ and their density $\mathcal{H}$, is given by
\begin{align}
H\equiv\int_{V}\mathcal{H}d^{3}x&=\int_{V} \psi^{\dagger}\Big(-\nabla^{2}+\frac{m^{2}c^{2}}{2\hbar^{2}}\Big)\psi d^{3}x\nonumber\\
&+\nabla\cdot\int_{V} \psi^{\dagger}\nabla\psi d^{3}x
\,
\end{align}
The second term is a divergence which does not change the action, and can be neglected. We collect the final expression
\begin{equation}\label{expectation_value}
H=\int_{V} \psi^{\dagger}\Big(-\nabla^{2}+\frac{m^{2}c^{2}}{2\hbar^{2}}\Big)\psi d^{3}x=\int_{V} \psi^{\dagger}\widehat{H}\psi d^{3}x=<\widehat{H}>
\end{equation}
\\
Since, $\widehat{p}=-i\hbar \nabla$, we can write the {\emph{canonical}} Hamiltonian operator $\hat{H_{c}}$ as
\begin{equation}\label{Hamiltonian operator}
\hat{H_{c}}=-\nabla^{2}+\frac{m^{2}c^{2}}{2\hbar^{2}}=\frac{\hat{p}^{2}}{\hbar^{2}}+ \frac{m^{2}c^{2}}{2\hbar^{2}}
\,
\end{equation}
\\
It is easy to see that the operator $\hat{H_{c}}$ is hermitian given their own definition, $\hat{H_{c}}^{\dagger}=\hat{H_{c}}$. On the other hand, the vanishing of the divergence, $\partial_{\mu}T^{\mu}_{\nu}=0$ implies, $d/dt(\int_{V} T^{0}_{0}d^{3}{\bf x})=d/dt(<\hat{H_{c}}>)=0$. Therefore, the expectation value (\ref{expectation_value}) turns out to be a constant of motion.

The Hamiltonian formalism allows us to reinterpret some of the results obtained previously. For instance, the conserved ``charge" (\ref{charge}) associated to the invariance of the Lagrangian under the global phase transformation (\ref{gauge}) can be written as
\begin{align}\label{electric_charge}
Q=\displaystyle\int_{V} J^{0}d^{3}{\bf x}&=\frac{i\theta}{2c}\int_{V} \Big[\Big(\boldsymbol{\sigma}\cdot\nabla\psi^{\dagger}\Big)\psi-\Big(\boldsymbol{\sigma}\cdot\nabla\psi\Big)\psi^{\dagger}\Big]d^{3}{\bf x}\nonumber\\
&=i\theta\int_{V} \Big(\pi(x)\psi(x)-\pi^{\dagger}(x)\psi^{\dagger}(x)\Big)d^{3}{\bf x}
\end{align}
Where we have employed the canonically conjugated momenta that were derived in (\ref{canonical_momenta}).
The physical interpretation of this formula is now transparent. The parameter $\theta$ is the electric charge times a constant. Therefore the conservation law (\ref{divergence}) is expressing nothing but the conservation of the electric charge. If $\psi(x)=\psi^{\dagger}(x)$ then according to (\ref{electric_charge}), $Q=0$ and the field describes a neutral particle. On the other hand, with this explicit expression for the charge given in terms of the fields and their canonical momenta, one is ready to make the next step and promote $Q$ from a classical quantity, to a quantum operator (the charge operator). In other words, the transition $Q\rightarrow\hat{Q}$ is automatic from (\ref{electric_charge}), once the standard anticommutation rules for a fermionic field, $\{\psi(x),\pi(y)\}=i\hbar\delta^{3}(x-y)$, are fixed. This task, however, will not be undertaken in this work.

\subsection{Canonical Hamiltonian Operator Vs Time Evolution Operator}
From the above equations, we can say some important things about the canonical Hamiltonian operator $\hat{H_{c}}$. In the first place it does not depend on time. Secondly, this canonical Hamiltonian involves certain time evolution, but {\emph{stricto sensu}}, it will not be equal to the pure time evolution operator, i.e, $\hat{H_{c}}\neq\mathcal{\hat{H}}$, where $\mathcal{H}$ is the standard Hamiltonian operator defined in (\ref{standard Hamiltonian}). To show this point, let us consider the action of the canonical Hamiltonian $\hat{H_{c}}$ over a plane wave solution of the wave equation

\begin{align}
\hat{H_{c}}\psi_{0}e^{i(\vec{k}\cdot\vec{r}-\omega t)}&=\Big(-\nabla^{2}+\frac{m^{2}c^{2}}{2\hbar^{2}}\Big)\psi_{0}e^{i(\vec{k}\cdot\vec{r}-\omega t)}\nonumber\\
&=\Big(k^{2}+\frac{m^{2}c^{2}}{2\hbar^{2}}\Big)\psi_{0}e^{i(\vec{k}\cdot\vec{r}-\omega t)}=\frac{\omega k}{c}\psi_{0}e^{i(\vec{k}\cdot\vec{r}-\omega t)}
\end{align}
where we have used the dispersion relation $\omega(k)$, associated to the positive energy solution, $\omega(k)=ck+\frac{m^{2}c^{3}}{2\hbar^{2}k}$. We have therefore obtained an eigenvalue equation, $\hat{H_{c}}\psi_{+}=\lambda_{+}\psi_{+}$, with $\lambda_{+}=\omega k/c$. Similarly, for the negative energy solution, $\psi_{-}=\boldsymbol\psi_{0}\exp(i(-\vec{k}\cdot\vec{r}+\omega t))$, one finds:
\small
\begin{align}
\hat{H_{c}}\psi_{0}e^{i(-\vec{k}\cdot\vec{r}+\omega t)}&=\Big(-\nabla^{2}+\frac{m^{2}c^{2}}{2\hbar^{2}}\Big)\psi_{0}e^{i(-\vec{k}\cdot\vec{r}+\omega t)}\nonumber\\
&=\Big(k^{2}+\frac{m^{2}c^{2}}{2\hbar^{2}}\Big)\psi_{0}e^{i(-\vec{k}\cdot\vec{r}+\omega t)}=-\frac{\omega k}{c}\psi_{0}e^{i(-\vec{k}\cdot\vec{r}+\omega t)}
\end{align}
\normalsize
These relations can be condensed in the compact expression, $\hat{H_{c}}\psi_{\pm}=\pm (\omega k/c)\psi_{\pm}$. Since the standard Hamiltonian verifies a different eigenvalue equation, namely, $\mathcal{\hat{H}}\psi_{\pm}=\pm\hbar\omega\psi_{\pm}$, it is obvious the non-equivalence of both operators. However, a close algebraic relation exists among them, and can be found paying attention to the structure of the wave equation. Indeed, since we already know that the wave equation (\ref{compact}) can be written in a Hamiltonian form (\ref{covariant}), we can identify
 \begin{equation}
(\boldsymbol{-i\sigma}\cdot\nabla)\mathcal{\hat{H}}\psi=c\hbar \Big(-\nabla^{2}+\frac{m^{2}c^{2}}{2\hbar^{2}}\Big)\psi=c\hbar\hat{H}_{c}\psi
\,
\end{equation}
Then, the exact algebraic relation between $\mathcal{\hat{H}}$ and $\hat{H}_{c}$ is given by the operator equation
\begin{equation}
(\boldsymbol{-i\sigma}\cdot\nabla)\mathcal{\hat{H}}=c\hbar \hat{H}_{c}
\,
\end{equation}
Another interesting consequence that can be extracted from the above relations is that, $[\mathcal{\hat{H}},\hat{H_{c}}]=0$. Of course this implies the conservation of $\hat{H}_{c}$, which is consistent with the result, $d/dt(<\hat{H_{c}}>)=0$, that we derived in the previous section from the canonical formalism. Then, the field theory developed here possesses all the ingredients required to proceed further. In particular, the canonical quantization and the construction of the Hilbert space of physical states are tasks that seem attainable once a Lagrangian formulation is provided. These important questions will be addressed in a forthcoming work.
\section{Discussion}
\label{conclusions}
\thispagestyle{empty}

\noindent
In this work we have presented a wave equation that works for particles whose energy can be approximated by, (ignoring constants), the relation $E\simeq p+\frac{m^{2}}{2p}$.
 If such energy-momentum relation encloses a ``hidden" wave equation, then this wave equation can only be the one that we have introduced in this paper, which is a hyperbolic second order linear PDE with well-behaved physical solutions. As we have demonstrated, it can be useful to explain some properties of ultra-relativistic particles. For instance, the family of plane-wave functions usually employed in the theory of neutrino oscillations (\ref{neutrino_waves2}) are only particular solutions of the wave equation discussed here.
Indeed, such as the Dirac equation, the wave equation (\ref{compact}) describes particles of spin $1/2$. In fact, the spin operators are incorporated in a natural way by means of the Pauli matrices, which emerge explicitly in the square root of the Laplacian that appears in the derivation of the wave equation. Interestingly enough, a similar wave equation (\ref{compact2}), can describe massive ultrarelativistic bosons of spin 1, if we replace the Pauli matrices $\boldsymbol{\sigma}$ (\ref{Pauli}), by the Majorana-Oppenheimer $\boldsymbol{\alpha}$ (\ref{Majorana-Oppenheimer}), maintaining the rest of the wave equation unaltered. Therefore the study of the behaviour of this wave equation under different interactions $V(x)$, will allow to enlarge the possible family of solutions, which may be useful to improve the understanding of ultra-relativistic processes, including perhaps the ultra-high energy cosmic-rays. 

In addition, a detailed Lagrangian formulation of the wave equation was also provided. In particular, we have proved that this is a consistent theory, where, through the symmetries of the Lagrangian density, some standard and well defined conservation laws are derived in a natural way. On the other hand, by means of the Hamiltonian formalism, we have demostrated that in the free case the expectation value of the operator, $-\nabla^{2}+m^{2}c^{2}/2\hbar^{2}$ is a conserved constant of motion (\ref{expectation_value}).\\
In conclusion, we point out that all the consistent matter wave equations in Physics derive of non-trivial energy-momentum relations. Indeed, if we think of the space of possible non-trivial energy-momentum configurations, we will realize that it is quite constrained: It seems that there are only four consistent possibilites: i). The non-relativistic, $E=p^{2}/2m$. ii). The linear, $E=\alpha^{i} p_{i}+\beta m$. iii). The quadratic, $E^{2}=p^{2}+m^{2}$. iv). The ultra-relativistic, $E=p+m^{2}/2p$. The first three options are all associated with consistent wave equations that describe particles with different properties in their appropriate physical regime. The study of option iv) deserves an analysis, and has been the subject of this work. The hypothetical existence of another matter wave equation is a very interesting possibility that deserves to receive further attention. In this sense, we point out that the field theory presented in this work is a natural alternative to the Dirac wave equation at very high energies. We have demostrated that it reproduces some standard results of the Dirac theory in the limit $p>>mc$, in a quite natural way (\ref{neutrino_waves2}). Besides, it incorporates the possibility of an explicit (and small) Lorentz invariance violation. However, our theory is far from being completely satisfactory. It lacks a canonical quantization, and the construction of a consistent Hilbert space; Nevertheless, with a Lagrangian formulation and a Hamiltonian formalism, the required ingredients to carry out these tasks are available.  

\end{document}